# Irrational Dynamical Variables
# and the Measurement Problem in Quantum Mechanics

Christopher Engelhardt*

8 July 2015

The quantum mechanical measurement process is considered. A hypothetical concept of irrational dynamical variables is proposed. A possible definition of measurement is discussed along with a mathematical method to calculate experimental result probabilities. The postulates of quantum mechanics are analyzed and modified. Thought experiments and implications are considered.

**1. History and Introduction**

The topics of this paper are several related ideas on the quantum mechanical measurement process and quantum superposition. First, a summary of key concepts in quantum mechanics and a brief introduction to the so-called measurement problem are given.

In late 1925, Erwin Schrödinger formulated the famous Schrödinger equation which governs nonrelativistic quantum mechanics.[1] This equation describes the evolution of a wave, however the exact nature of this wave remains unclear. In 1926, Max Born formulated the Born rule that states the wave function is a probability amplitude and describes how to calculate probabilities for experimental results.[2] The Born rule introduced probabilistic elements to Schrödinger's deterministic equation and thus conflict. Specifically, the conflict arises during the measurement process.

Quantum mechanics, quantum states, and the measurement process are frequently formulated using linear algebra. Suppose an elementary particle is described by the state $|\psi\rangle$ and a macroscopic measuring device is described by the state $|\phi\rangle$. If the elementary particle is initially in an eigenstate, $|\psi_n\rangle$, of the corresponding observable before a measurement, the measurement process will transform the device state from the general state, $|\phi\rangle$, to the registered state, $|\phi_n\rangle$. The particle state, $|\psi_n\rangle$, will remain unchanged. This is shown in Process (1.1). Likewise, any other eigenstate, say $|\psi_m\rangle$ in Process (1.2), will experience the same behavior.

$$|\psi_n\rangle|\phi\rangle \rightarrow |\psi_n\rangle|\phi_n\rangle \quad (1.1)$$

$$|\psi_m\rangle|\phi\rangle \rightarrow |\psi_m\rangle|\phi_m\rangle \quad (1.2)$$

Because the Schrödinger equation is linear, superpositions of solutions should also be solutions. Thus, one can construct a state that is a superposition of both $|\psi_n\rangle$ and $|\psi_m\rangle$. When this state is measured, the linear Schrödinger equation dictates that the resulting measurement device would also be a superposition as in Process (1.3). However, this is never observed experimentally. What is observed experimentally is described by von Neumann's reduction postulate.[3] The von Neumann reduction postulate describes the measurement process when the elementary particle is not initially in an eigenstate. In this scenario, a measurement causes a nondeterministic and nonlinear reduction in the state vector producing $|\psi_n\rangle$ in certain experiments, Process (1.4), and $|\psi_m\rangle$ in others, Process (1.5). Transition from (1.3) to one of either (1.4) or (1.5) is referred to as the strong von Neumann projection, state vector reduction, or wave function collapse. Superpositions are expected from the time evolution of the linear Schrödinger equation, however they are never observed. This is the source of the measurement problem.

---

*chrisengelhardt@mykolab.com

$$c_n|\psi_n\rangle|\phi\rangle + c_m|\psi_m\rangle|\phi\rangle \rightarrow c_n|\psi_n\rangle|\phi_n\rangle + c_m|\psi_m\rangle|\phi_m\rangle \tag{1.3}$$

$$c_n|\psi_n\rangle|\phi_n\rangle + c_m|\psi_m\rangle|\phi_m\rangle \rightarrow |\psi_n\rangle|\phi_n\rangle \tag{1.4}$$

$$c_n|\psi_n\rangle|\phi_n\rangle + c_m|\psi_m\rangle|\phi_m\rangle \rightarrow |\psi_m\rangle|\phi_m\rangle \tag{1.5}$$

Continuing the mathematical formalism, the Born rule states that the probability of an experimental outcome is $|c_n|^2$ for Process (1.4) and $|c_m|^2$ for Process (1.5). To constrain the total probability to unity, Equation (1.6) is required.

$$\sum_i |c_i|^2 = 1 \tag{1.6}$$

The complete description of state vector reduction, which includes degenerate measurements, is shown by Process (1.7). Here, $P_\omega$ is the projection operator for an eigenstate or eigenspace with eigenvalue $\omega$. If the quantum state is measured and the eigenvalue $\omega$ is obtained, the state will collapse to $|\omega\rangle$ if $\omega$ is not degenerate. If $\omega$ is degenerate, the state will collapse to the eigenspace comprised of all degenerate $|\omega\rangle$ states.[4]

$$|\psi\rangle \rightarrow \frac{P_\omega|\psi\rangle}{\langle P_\omega\psi|P_\omega\psi\rangle^{1/2}} \tag{1.7}$$

With this background, the current postulates of nonrelativistic quantum mechanics for one degree of freedom under ideal measurement can be presented. The following form is reproduced from Shankar, however other authors organize the postulates in somewhat different ways.[4]

I. Particles are represented by state vectors in Hilbert spaces.

II. Position and momentum observables are represented by the Hermitian operators $X$ and $P$ with matrix elements given in Equations (1.8) and (1.9). Additional dynamical variables are obtained by substituting $X$ into classical position variables and $P$ into classical momentum variables.

$$\langle x|X|x'\rangle = x\delta(x-x') \tag{1.8}$$

$$\langle x|P|x'\rangle = -i\hbar\delta'(x-x') \tag{1.9}$$

III. Measurements will result in an eigenvalue of the observable with a probability given by the Born rule, Equation (1.10). The object's state vector will collapse to the corresponding eigenstate or eigenspace.

$$P(\omega) = |\langle\omega|\psi\rangle|^2 \tag{1.10}$$

IV. The state vector evolves according to the Schrödinger equation, Equation (1.11), with an appropriate Hamiltonian, $H$.

$$i\hbar\frac{d}{dt}|\psi(t)\rangle = H|\psi(t)\rangle \tag{1.11}$$

Next, a brief review of theoretical measurements is given. Certain position measurements can be approximated by the hypothetical Heisenberg microscope. In this thought experiment, photons are used to measure the position of an electron under a microscope lens. In order to ideally measure the electron position, infinitely high momentum photons must be used. These high momentum photons produce an indeterminism in the momentum of the electron being measured.[5]

Similar concepts are found in hypothetical momentum measurements. One way to perform an ideal momentum measurement is by Compton scattering. In this method, a photon strikes a particle and loses energy thus changing the photon's wavelength. The wavelength is measured before and after the collision. By applying energy and momentum conservation to the system, the momentum of the particle is calculated. It can be shown that this method requires photons of infinitesimally low momentum to not alter the momentum of the particle but produces an indeterminism in position. This measurement is considered ideal.[6]

Measurement devices may interact with a quantum system directly or indirectly. In a direct measurement, the quantum system interacts with only the measuring device. In an indirect measurement, a probe is first correlated to the quantum system. In a second step, the probe interacts with the measuring device. The wave function of both the probe and quantum system are reduced in the second step.[7]

The above analysis assumes precise measurements. Realistic measurements can be described by imprecise measurement theory. To do so, the resolution amplitude, $\Upsilon_{\bar{a}a}$, is introduced. $\Upsilon_{\bar{a}a}$ is defined as the amplitude to obtain a measurement result of $\bar{a}$ given that the true eigenvalue is $a$ after a measurement of $A$. The resolution amplitude is entirely dependent on the specific experimental measuring device used. Two equations succinctly define imprecise measurement theory. First, Equation (1.12) gives the probability of an experiment returning a measured value $\bar{a}$. Second, after $\bar{a}$ is returned, the wave function of the measured object changes to the state shown in Process (1.13). Resolution amplitudes also satisfy a normalization condition, Equation (1.14).[8, 9]

$$P(\bar{a}) = \sum_a |\Upsilon_{\bar{a}a}|^2 |\langle a|\psi\rangle|^2 \tag{1.12}$$

$$|\psi\rangle \to \sum_a |a\rangle \frac{\Upsilon_{\bar{a}a}\langle a|\psi\rangle}{\sqrt{P(\bar{a})}} \tag{1.13}$$

$$\sum_{\bar{a}} |\Upsilon_{\bar{a}a}|^2 = 1 \tag{1.14}$$

Gardiner and Zoller discuss important special cases of imprecise measurement. One is that precise measurement theory is recovered from imprecise measurement theory when resolution amplitudes are equal to unit modulus for $\bar{a}=a$ and zero for $\bar{a} \neq a$. Another special case defines the reduced operator, $\tilde{A}$. If the resolution amplitudes are equal to zero or one and if each $\bar{a}$ represents a unique range of $a$ values, then the measurement can be considered that of the reduced operator. $\tilde{A}$ is given in Equation (1.15). $\hat{\Upsilon}_{\bar{a}}$ is defined in Equation (1.16).[9]

$$\tilde{A} = \sum_{\bar{a}} \hat{\Upsilon}_{\bar{a}} \tag{1.15}$$

$$\hat{\Upsilon}_{\bar{a}} \equiv \sum_a \Upsilon_{\bar{a}a} |a\rangle\langle a| \tag{1.16}$$

Several interpretations have been devised to explain the conceptual conflicts of quantum theory and the measurement problem in particular. The Copenhagen interpretation is in agreement with the postulates as they are currently presented.[10] Popular interpretations include information-based and Everett's many-worlds amongst others.[11, 12] There is no current consensus.[10, 13] The concept of decoherence has been proposed as a solution to the measurement problem. Decoherence claims that device eigenstates are preferred due to the macroscopic nature of any measuring device and its inability to be isolated from the environment.[14, 15] However, decoherence is not without its critics who claim that decoherence is governed by the Schrödinger equation while measurement uniqueness is beyond the equation.[10]

In the following, a hypothetical mechanism that constitutes quantum measurement is proposed. Consequently, the application of the Born rule and the concept of dynamical variables are generalized. Thus, postulates II and III are modified.

**2. Proposed Theory**

The Heisenberg microscope and Compton scattering, although insightful, do not define the measurement process. In addition, both processes can be considered incomplete measurements. Considering a position measurement using the Heisenberg microscope, the photon can be viewed as an intermediate particle; the photon's position must still be measured. Considering a momentum measurement by Compton scattering, the recoiled photon may again be viewed as an intermediate particle; a measurement of its wavelength is still required. Both thought experiments push the fundamental measurement problem aside.

The hypothesis considered here is that an actual measuring device is a set of particles with quantum states specifically related to the measurement space of interest. For example, a fluorescent screen which performs position measurements on electrons is composed of atoms that are localized in position space. The equivalent device for a momentum measurement would be a set of particles localized in momentum space.

Consider a series of overlapping beams of particles as shown in Figure (2.1). Say several machines each emit particles at a narrowly defined wavelength. The wave functions are highly localized in momentum space and highly spread in position space. An electron traveling through the beams will have its wave function altered. Presumably, the electron wave function will become localized in momentum space; it can be considered that a momentum measurement has occurred. This setup should be considered a direct momentum measurement as opposed to Compton scattering. Although not required, the resulting momentum value can be inferred from a change in the intensity of one specific beam.

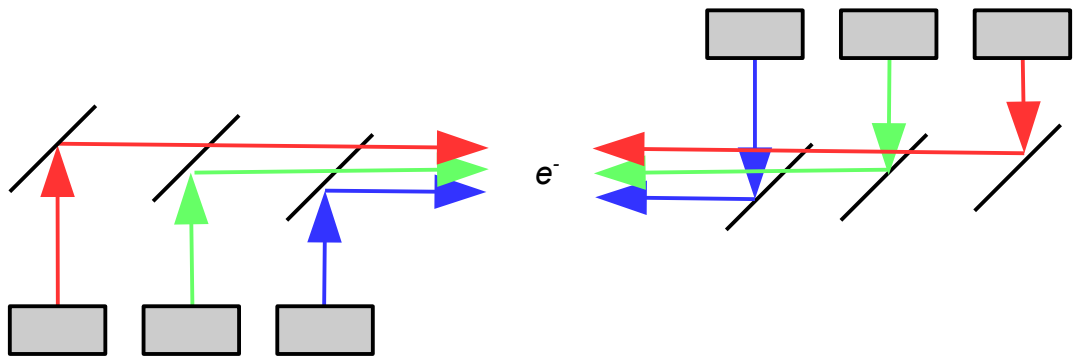

Figure (2.1): Direct Momentum Measurement Device

This method may be applied to measurements of other dynamical variables. For example, a direct energy measurement device should be comprised of particles with wave functions localized in energy space. Moreover, this method may also be generalized to spaces that have no classical equivalent. In classical physics, all dynamical variables are comprised of position and momentum. The concept of measurement spaces that cannot be comprised of position and momentum, call them irrational spaces, has no meaning in classical physics. An arbitrary space between position and momentum in quantum mechanics is well defined by a set of orthogonal states. Conceivably, particles in any orthogonal set of states can act as a measurement device, and the dynamical variable they are measuring can be considered an irrational dynamical variable.

This concept can be extended further to systems with nonorthogonal states. However, for nonorthogonal systems the Born rule is insufficient. Instead, probabilities should be calculated by considering the cumulative measurement to be individual measurements performed in succession. If the device is assumed perfectly ideal, every permutation of all measurement states, $|a\rangle$, should be included and weighted equally. Figure (2.2) illustrates a convenient method to calculate probabilities on an example two state system. Here, there are two permutations; $|a_1\rangle$, $|a_2\rangle$ is represented by the top set, and $|a_2\rangle$, $|a_1\rangle$ is represented by the bottom set.

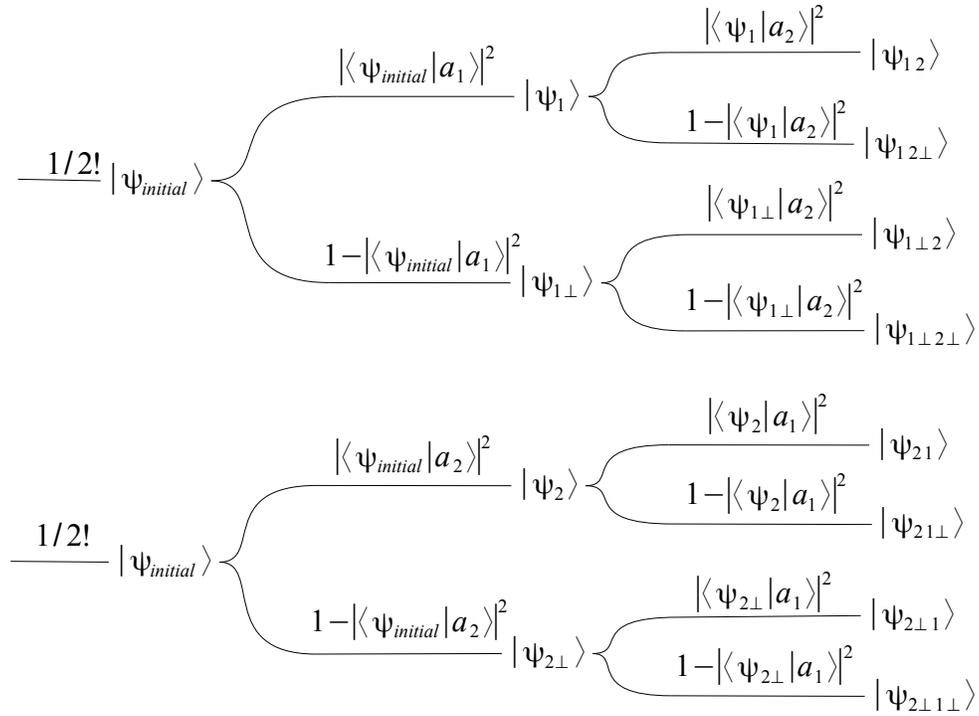

Figure (2.2): Measurement Permutations with the Born Rule

Considering the top set in Figure (2.2), $|\psi_{initial}\rangle$ is first measured by $|a_1\rangle$ producing two possible alternatives. Traversing the upper fork is the alternative that $|\psi_{initial}\rangle$ is projected onto $|a_1\rangle$. Traversing the bottom fork is the alternative that $|\psi_{initial}\rangle$ is projected onto the space orthogonal to $|a_1\rangle$. Both of these are equivalent to what would be expected from the Born rule for a reduced operator. Each branch now forks again as $|a_2\rangle$ is measured. A projection onto $|a_2\rangle$ or onto the space orthogonal to $|a_2\rangle$ occurs with different probabilities. The analogous process occurs for the second permutation, $|a_2\rangle$, $|a_1\rangle$, shown on the bottom set in Figure (2.2).

There are now eight possible $|\psi_{final}\rangle$ states and eight probabilities at the conclusion of the measurement. To calculate the probability of any $|\psi_{final}\rangle$, simply multiply the probabilities that produced the state including the initial permutation factor. To determine the measurement results, select all affirmative $a$ values in the path.

Under this framework, postulates II and III require modification. Postulate II should include nonorthogonal and irrational dynamical variables and declare that observables must be fully defined by all $|a\rangle$ vectors comprising the measurement space. $X$, $P$, and derived classical dynamical variables may be considered special cases of this more general definition. The Born rule should also be generalized in postulate III to include nonorthogonal measurement devices by calculating probabilities in succession. For orthogonal measurement spaces, postulate III reduces to the single Born rule.

II. Dynamical variables are defined by a set of measurement states.

III. Probabilities are calculated by applying the Born rule to each measurement state. The object's state vector will collapse to the corresponding measurement states or null result spaces.

## 3. Example Calculations

In this section, example probability calculations are performed using the procedure shown in Figure (2.2) as opposed to the single Born rule. In all of the following figures, $|\psi_{initial}\rangle$ is represented by the black vector, $|a\rangle$ states are represented by the colored vectors, and possible $|\psi_{final}\rangle$ states are represented by the grey vectors. In all figures, $|\psi_{initial}\rangle$ is equal to $\sqrt{3}/2|x\rangle + 1/2|y\rangle$. Also, note that all vectors should be normalized, however some have been arbitrarily shortened or lengthened for visual clarity with no physical significance.

Example 3.1. Orthogonal Measurement Device
The following result illustrates that the suggested calculation method reproduces Born rule probabilities for the simple two state system with orthogonal $|a\rangle$ states. The resultant states and probabilities are as expected.

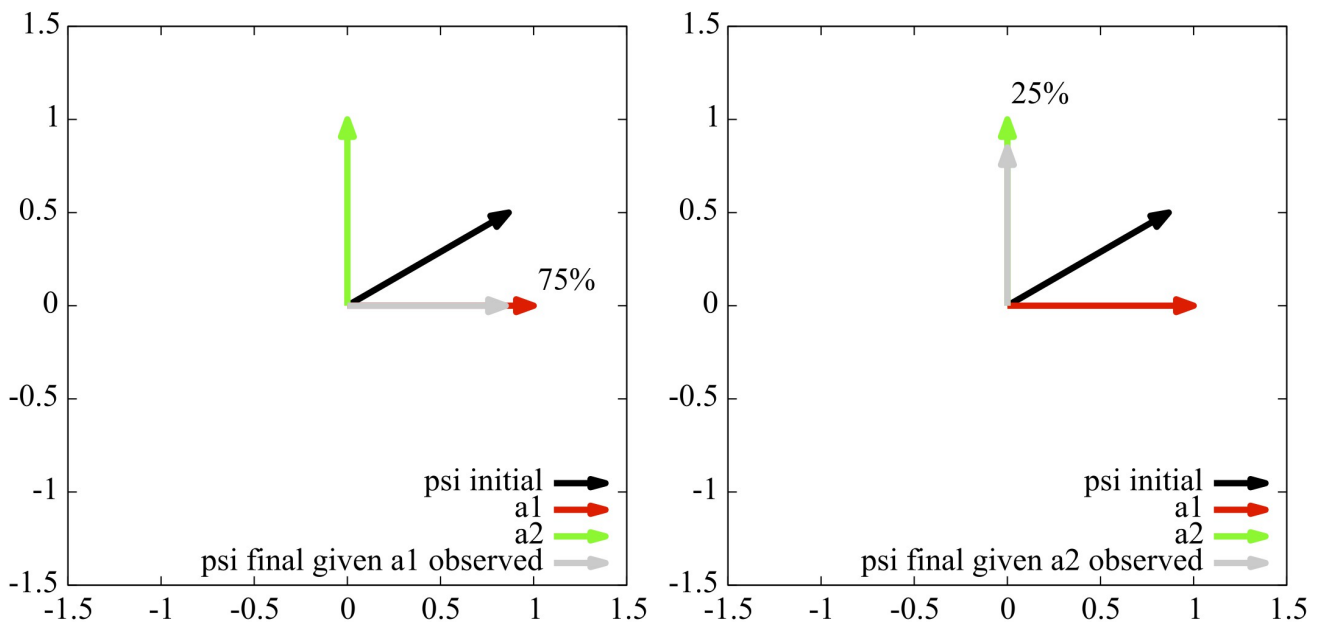

Figures (3.1.1) and (3.1.2): Orthogonal Measurement

Example 3.2. Repeated Orthogonal Measurement Device

Figures (3.2.1) and (3.2.2) show results when an extra measurement state is added on top of a previous one. The only change as compared to Example 3.1 is that there are two simultaneous results.

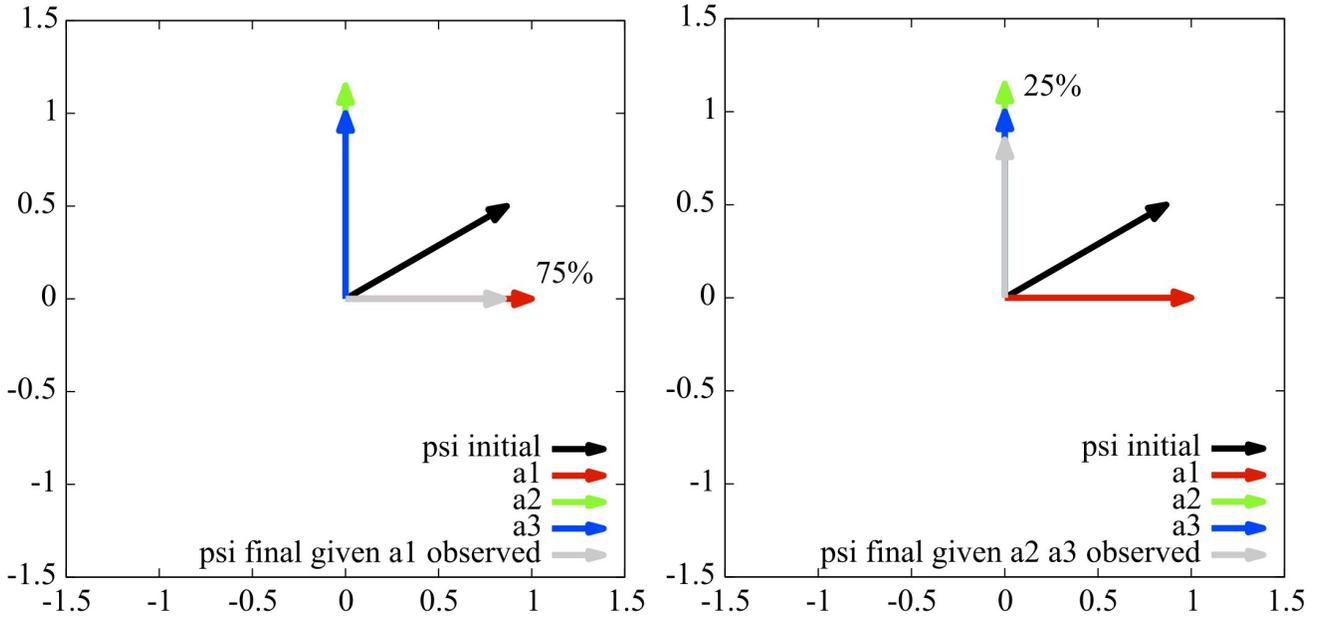

Figures (3.2.1) and (3.2.2): Orthogonal Measurement with Repeated Measurement States

Example 3.3. Nonorthogonal Measurement Device

Example 3.1 is altered to produce nonorthogonal measurement states by rotating $|a_2\rangle$ 10 degrees. When *a1* is observed, there are two possible $|\psi_{final}\rangle$ states. Only one state overlaps $|a_1\rangle$. 36.37% of the measurements will return a value of *a1* with $|\psi_{final}\rangle$ orthogonal to $|a_2\rangle$. Physically, this corresponds to first an affirmative measurement of $|a_1\rangle$ followed by a null measurement of $|a_2\rangle$. An analogous process occurs for the returned value of *a2* which can be seen in Figure (3.3.2). Figure (3.3.3) shows multiple affirmative results, and Figure (3.3.4) shows no results.

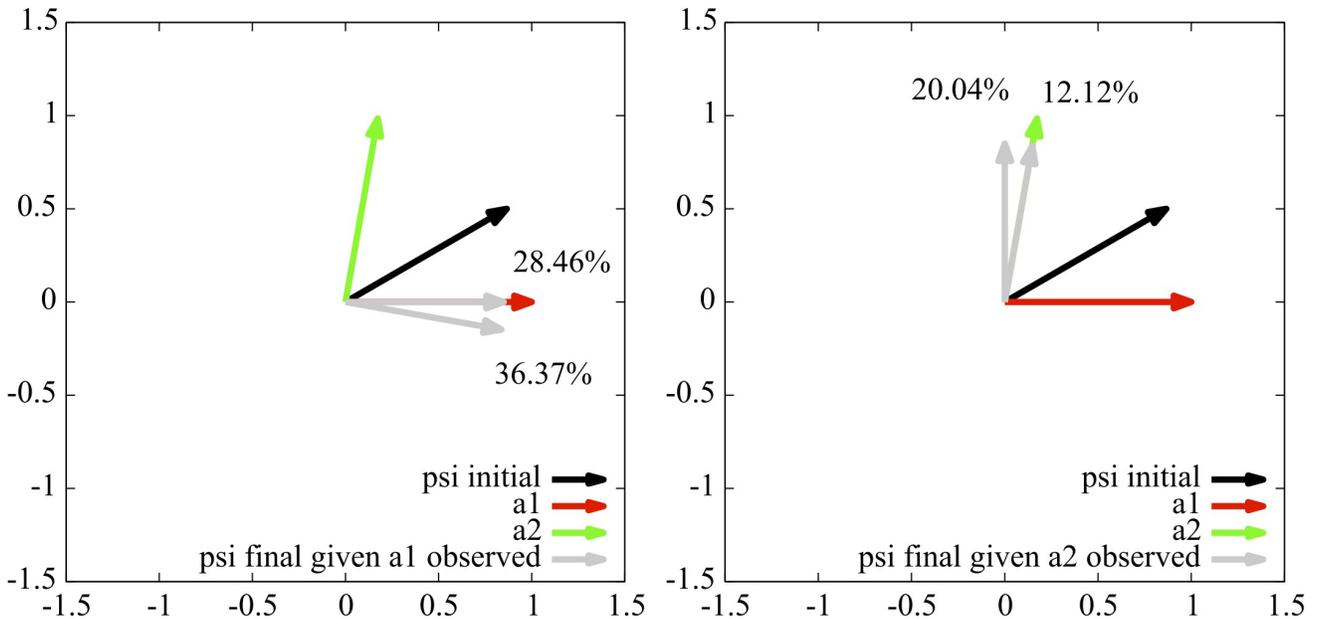

Figures (3.3.1) and (3.3.2): Nonorthogonal Measurement

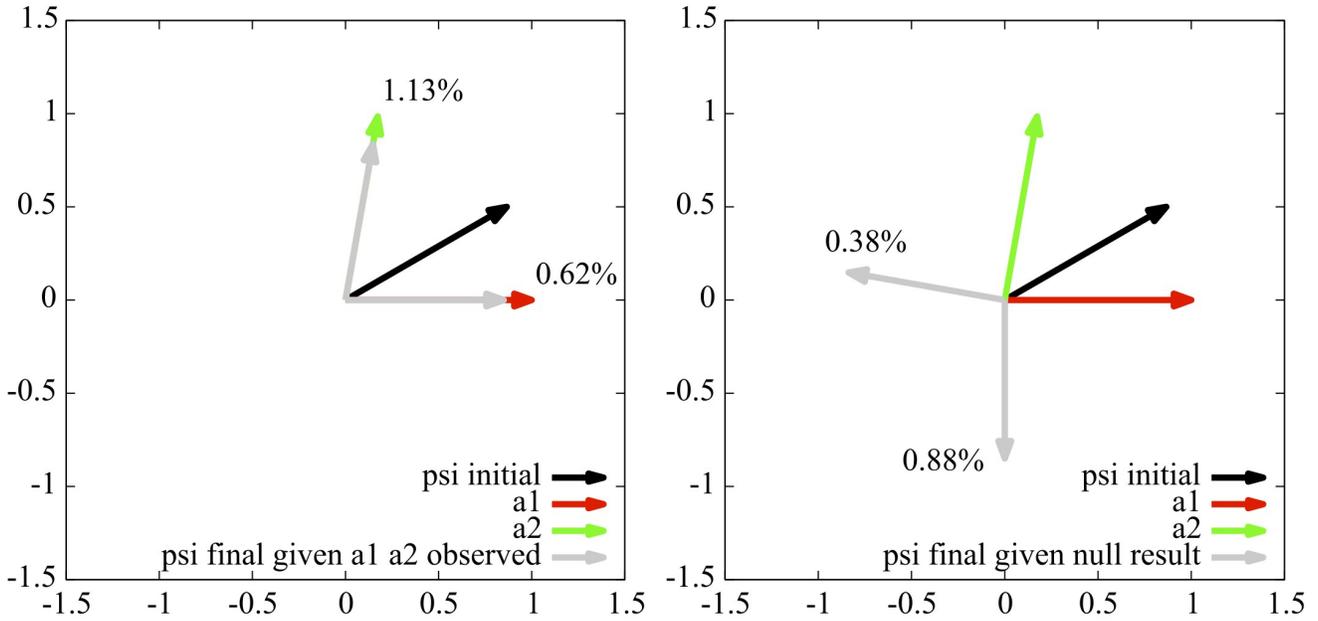

Figures (3.3.3) and (3.3.4): Nonorthogonal Measurement

### Example 3.4. Repeated Nonorthogonal Measurement Device

Example 3.3 is modified to include an additional overlapping measurement state. Only two selected example results are presented here. Comparing Figure (3.4.1) to Figure (3.3.1), it is shown that adding overlapping measurement states to nonorthogonal measurements changes both the total probability of observing *a1* and the resulting individual final probabilities. This is in contrast to the orthogonal case.

Another interesting point is that for overlapping nonorthogonal measurements not all overlapping states may display the same result. As shown in Figure (3.4.2), *a3* is measured affirmative while the overlapping *a2* is null, a weaker result. Also, the resultant object states are very different than the registered measurement state.

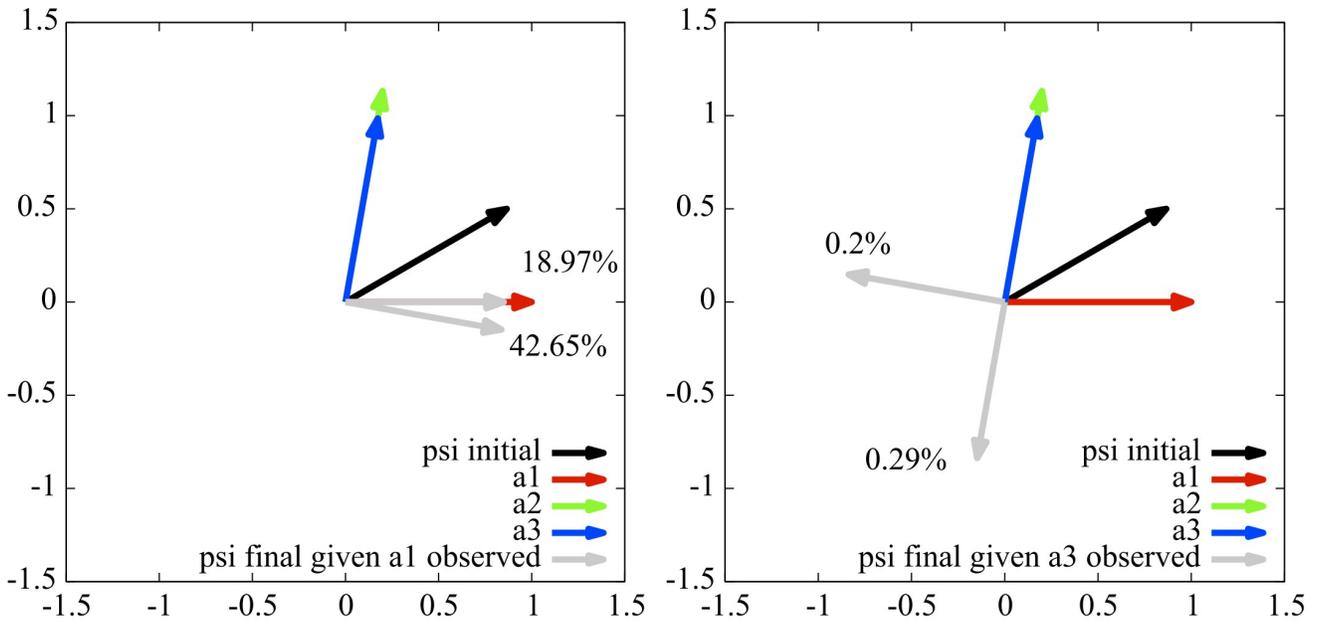

Figures (3.4.1) and (3.4.2): Selected Examples of Nonorthogonal Measurement with Repeated Measurement States

Example 3.5. Random Measurement Device

As a final example, consider the results as measurement states are added randomly. Multiple results may be returned, and the final object state has many possibilities. This situation more closely approximates what is traditionally considered a non-measurement among interacting particles.

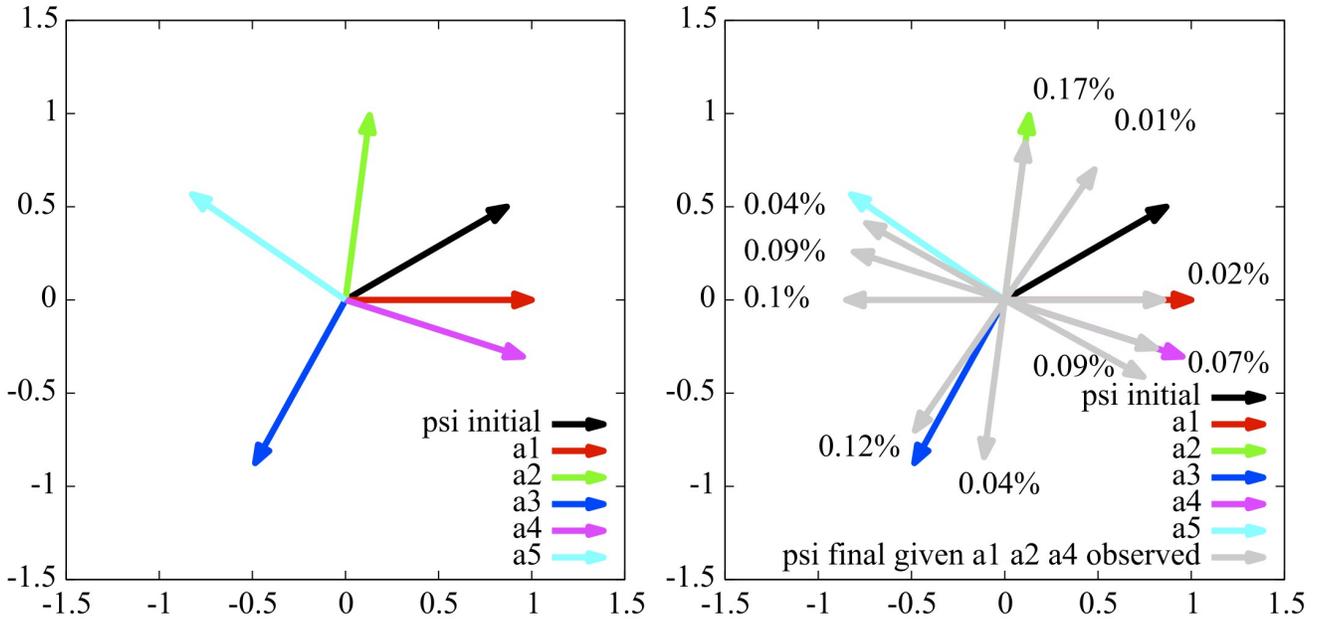

Figures (3.5.1) and (3.5.2): Initial Configuration and a Selected Example of Measurement
with Nonorthogonal and Random Measurement States

**4. Implications**

An important implication is that a dynamical variable should be viewed as a set of possible measurement states. These sets can be orthogonal or nonorthogonal. In the orthogonal case, the Born rule need only be applied once to calculate probabilities, and multiple results will never be observed. The orthogonal case may include irrational dynamical variables that do not exist in classical physics, although these are just as valid as position or momentum variables. Consequently, for every $|\psi\rangle$, there exists an irrational dynamical variable where $|\psi\rangle$ is a measurable eigenvector.

Nonorthogonal sets of measurement states may return a single result, multiple results, or no result. A measurement of nonorthogonal sets may be interpreted as simultaneous measurements of multiple orthogonal irrational dynamical variables, or it may be interpreted as a measurement of one more complicated dynamical variable. Error or uncertainty are present.

The concepts presented may be used to understand macroscopic measurement. Consider a measuring device composed of a single particle with only two possible measurement results: an affirmative result or a null result. In the affirmative case, the wave function of the measured object projects to $|a\rangle$. In the null case, the wave function projects to the space orthogonal to $|a\rangle$. Essentially, the single particle is measuring the reduced operator of a, generally irrational, dynamical variable. Now imagine adding a second particle. It can be considered that the device is now made of two particles measuring a more detailed reduced operator. It can also be considered that there are two devices measuring different dynamical variables simultaneously. A third view would be taking the limit as two spatially distinct measuring particles are brought close together. The thought experiment of adding particles can be continued up to the macroscopic level.

Traditional measurement devices, which are more technologically elaborate, may be explained under this framework by considering stages of probes. For example, any measurement that involves a human requires multiple probe stages: a machine measures a quantum object, the human eye measures the light from a computer monitor, the brain measures information from the optic nerve, etc. Certain stages are direct measurements, and others are indirect measurements, i.e., they cause a collapse at the second stage. However, some stages are actually measurements of a different dynamical variable than desired and a calculation of the variable of interest. For example, a common momentum measurement is actually two position measurements and a calculation performed by the scientist. In the case of Compton scattering, a momentum measurement consists of two wavelength measurements and a calculation. Measuring an alternative dynamical variable may have practical advantages, but according to the proposed theory it is still a workaround.

Resolution amplitudes may also be used to understand the range of interactions between non-interacting particles and measured particles. In the general case, resolution amplitudes vary between zero and unity; measurement devices do not project wave functions perfectly, and measurement results are never without some error. Presumably, these resolution amplitudes can be calculated for a single particle measuring another single particle. These calculations depend on the type of particle interaction and are time dependent. In principle, the degree of orthogonality of resolution amplitudes can be used as a metric to judge the degree of measurement which includes the amount of error. Strong particle interaction implies highly orthogonal resolution amplitudes; weak particle interaction implies overlapping resolution amplitudes and thus highly error prone results.

An important note is that measurement states, $|a\rangle$, although specifically related, are not identical to device states, $|d\rangle$. $|d\rangle$ is the state of a particle in the measurement apparatus. $|a\rangle$ is the state that the measured object collapses to. The specific interaction between device particle and measured particle may make $|a\rangle$ and $|d\rangle$ very similar or very different. In the previous sections, this distinction has been neglected. Calculating $|a\rangle$ from $|d\rangle$ for a given particle interaction should be possible but is beyond the scope of this paper.

Experimentally, these concepts allow additional manipulation of wave functions. It is possible to prepare wave functions with more elaborate interference patterns using irrational and nonorthogonal measurement states. It would also be interesting to place several measurement devices of deliberately chosen irrational variables in succession to show a repeatedly measured wave function obeying perfect Schrödinger evolution.

## 5. Additional Thoughts

It may be possible to find experimental evidence of natural measurements which are conducted by ordered $|a\rangle$ states. For example, consider Planck's postulate of black-body radiation. Resonator energy is quantized naturally without any laboratory measurement device in use.[16, 17] It is possible that each particle is continually measuring every other particle, and the measurement states quickly rearrange to an ordered, discrete, steady pattern. Randomly perturbed states may revert into ordered states through this natural measurement system. Being a steady solution, the eigenvectors remain constant in time and represent energy of definite value. It is also possible that other phenomena, including interaction laws, may become simpler in terms of certain irrational dynamical variables.

It is interesting to consider the Schrödinger cat paradox in the context of the proposed theory.[18] The Schrödinger cat is essentially a macroscopic version of Equations (1.3) through (1.5). An important difference is that "alive" and "dead" are poorly defined macroscopic qualities as opposed to eigenstates or eigenvalues. Nevertheless, consider a cat that was somehow placed in a superposition of alive and dead states. There should be an irrational dynamical variable where this state is an eigenstate. From a practical perspective, measuring any state would be nearly impossible as the wave function of a macroscopic object exists in a configuration space with tremendously many dimensions. However, in principle, the irrational dynamical variable should exist, and there should be a device capable of measuring it.

Bell proved that local hidden variable theories are incompatible with quantum mechanics.[19] However, his proof does not apply to nonlocal hidden variable theories. It is possible that the full state and relative phase of the device constitute the hidden variable. The state and relative phase of the measuring device in the laboratory are never actually known; only the general macrostate is known. Perhaps the Born rule can be derived from considering all possible initial device states and relative phases. An interesting experiment would be to create a system with an elementary particle, a micro measurement device, and a macro measurement device. In this system, a scientist would be capable of measuring the measuring tool in addition to the elementary particle. It would be very interesting if the elementary particle's results were strongly correlated to the full initial state of the micro device.

## 6. Conclusions

This paper has discussed a hypothetical explanation of the measurement problem in quantum mechanics. Two slight modifications to the postulates of quantum mechanics have been proposed that may rectify the apparent conflict. First, dynamical variables must be described by their measurement states and associated result values. Second, the Born rule must be applied in succession to all measurement states. Qualitatively, the concept of a dynamical variable must be generalized to include the nonorthogonal and irrational. Resolution amplitudes and imprecise measurement theory may be used to explain the appearance of a distinction between Schrödinger evolution and laboratory measurements. Experimental measurement devices may be constructed by filling the desired measurement space with device particles of appropriate measurement states.